\documentstyle[]{laa}
\def\vect{\mathaccent"017E }

\newcommand{\be}{\begin{equation}}
\newcommand{\ee}{\end{equation}}
\begin{document}
%
\thesaurus{02(03.12.1; 07.19.1; 14.03.1; 18.02.1; 20.01.2)}
\title{Optimal reconstruction of the velocity and density field:
{\sc Potent} and max-flow algorithms}
%
\author{J.F.L. Simmons\inst{1}
  \and  A. Newsam\inst{1}
  \and  M.A. Hendry\inst{2}}
\offprints{J.F.L. Simmons} 
\institute{Department of Physics and Astronomy, 
	   University of Glasgow, Glasgow, UK
  \and Department of Astronomy, University of Sussex, Falmer, Brighton, UK}
\date{Accepted 8 April 1994}
%
\maketitle
\begin{abstract}
Although {\sc Potent} purports to use only radial velocities in reconstructing
the potential velocity field of galaxies, the derivation of transverse
components is implicit in the smoothing procedures adopted. Thus the possibility
arises of using nonradial line integrals to derive a smoothed velocity field.
For an inhomogeneous galaxy distribution the optimal path for integration need
not be radial, and can be obtained using max-flow algorithms. In this paper
we describe how one may use Dijkstra's algorithm to obtain this optimal path
and velocity field, and present the results of applying the algorithm to a 
realistic spatial distribution of galaxies. These results show that the method
has limited effect due to the large smoothing scales employed in 
{\sc Potent}. However, the viability of the technique is demonstrated and,
finally, we discuss other possible 
methods involving averaging over an ensemble of non-radial paths for improving 
a potential velocity field derived from redshifts.

\keywords{cosmology -- galaxies: redshifts of -- numerical 
methods -- radial velocities -- universe: structure of}
\end{abstract}
%
%
\section{Introduction}
One of the most effective methods for reconstructing peculiar velocity
fields of galaxies on the scale of 100 Mpc from observed redshifts has
been {\sc Potent} (Bertschinger and Dekel, \cite{Bert+89}, 
Bertschinger et al., \cite{Bert+90}, 
Dekel et al., \cite{Dekel+90} hereafter DBF, 
Dekel et al., \cite{Dekel+93}). 
The main difficulty in implementing this method is the sparseness of the data, 
as sky 
coverage of even the most complete redshift surveys is still highly 
anisotropic. Some of this incompleteness will be overcome by more rigorous 
and deeper surveys. However, the intrinsic distribution and clustering of 
galaxies and voids will always be a barrier to determining peculiar velocity 
fields accurately. 
In these circumstances it is important to
find the optimal way of reconstructing the peculiar velocity 
field. This in turn could lead us to a better understanding of 
large scale density inhomogeneities, and place improved constraints on the 
value of $\Omega$ and the various structure formation scenarios. 

The {\sc Potent} reconstruction technique has usually been presented as a 
method for obtaining the smoothed peculiar velocity field almost directly 
from observations of redshifts and distances of galaxies. Such an attractive 
idea often obscures the actual practice by which the smoothed potential 
field is constructed. In this paper we present a method
for obtaining the velocity potential from {\em non-radial\ }paths.  
This notion of using non-radial paths at first seems counter intuitive. Of 
course the observed redshifts and distances of galaxies at a set of points 
enables one to trivially construct a potential along the rays connecting the 
origin to those points. However, such a procedure will probably yield a noisy potential 
function which is obviously not defined in between rays. It is essentially 
the smoothing procedure used by {\sc Potent} which allows one to reconstruct 
the potential velocity field everywhere. {\sc Potent} uses a tensor window 
function, essentially a weighting function, to construct an 
{\em initial\ }smoothed vector velocity field which is not in the first 
instance potential. 
Line integration of this initial velocity field need not be along radial 
paths. One would expect the choice of sets of paths to affect the resulting 
potential velocity field obtained. 

The {\sc Potent} reconstruction is plagued by the problem of uncertainties 
and bias in the distance estimates of galaxies, and this problem enters 
primarily at the stage of calculating the peculiar radial velocity of a 
galaxy. In effect {\sc Potent} has always taken a window function with a 
large effective diameter (in the region of 2400kms$^{-1}$) compared with 
the region of interest out to about 6000kms$^{-1}$, which in itself 
demonstrates the importance of the choice of smoothing procedure on the 
reconstructed potential field. Since one is by no means confined to using 
radial path integrals, the possibility opens up of choosing paths that pass 
through regions where the number density of galaxies is high, and where the 
errors of distance estimation are low. Such 
paths should yield a more accurate potential velocity field.  Indeed we could
move away entirely from the ad hoc choice of window function, and attempt to
obtain a potential velocity field by averaging integrals taken over an 
ensemble of paths to any point. This idea, reminiscent of Feynman's path 
integral approach, might provide a powerful alternative to conventional 
{\sc Potent}, although we shall not discuss it in detail in this paper.

The structure of this paper is as follows. In 
Sect.~\ref{sec:orthPot} we outline the 
main steps used in {\sc Potent} to construct the potential velocity field, 
emphasising the aspects of the analysis that are most important to our new
approach. In Sect.~\ref{sec:nonrad} we discuss the method by which {\sc Potent} constructs 
an initial vectorial velocity field, and consider the covariance matrix of the 
errors on this field. In Sect.~\ref{sec:optpath} we consider the question of optimal paths 
of integration for the reconstruction of the velocity potential and in 
Sect.~\ref{sec:dijkstra} introduce a finite methods algorithm, Dijkstra's Algorithm, that 
can be used to obtain the path that minimises the integrated error on 
the velocity potential. In Sect.~\ref{sec:dijkandPot} we apply our new method to a spatial 
distribution of galaxies obtained from the Mathewson et al. (\cite{Mathew+92}) and 
Burstein et al. (\cite{Burstein+87}) galaxy surveys which have both redshifts
and redshift independent distance estimates for each galaxy. Finally in 
Sect.~\ref{sec:conclusion} we discuss our 
results and other possible methods of improving them.

\section{Conventional Potent}\label{sec:orthPot}

To understand our new approach to constructing potential velocity fields from
redshifts, we shall have to outline the main steps of the {\sc Potent} 
method of DBF. The essential idea of {\sc Potent} is to obtain the 
velocity potential of the peculiar velocity field by taking a radial line
integral of the observed quantities. Thus if we write the radial component of 
the peculiar velocity as $v = z-H r$
then the the velocity potential, $\Phi (\vect r)$, at position $\vect r$ may be 
written as
\be
\Phi (\vect r) = \int_0^{\vect r} v(s) ds
\ee
The evident difficulties in applying this equation arise from the fact that the 
galaxies are sampled at discrete points, and furthermore are obviously not 
located along rays, and that the distances to galaxies are only estimated
distances and hence will be subject to errors and possible biases.

The main steps of {\sc Potent} may be summarised as the following.
\begin{description}
\item[1.] Take  the measured redshift and distance, and hence infer
the peculiar velocity, of each individual galaxy. The distance can be 
estimated by a number of different methods, but will usually be obtained by 
some variant of the Tully Fisher or $D_n$-$\sigma$ relation. 
\item[2.] Construct an initial smoothed velocity field $\vect v$ at 
every point, $\vect r$,  by using a tensor window
function. This is achieved by minimising the weighted sum of square differences 
between the radial component of the
peculiar velocity and some simple velocity field model
(see Sect.~\ref{sec:nonrad}).  
The window function is usually taken to be gaussian, and
in the analyses of DBF it has a wide window of diameter approximately one 
third of the range of the survey. This initial smoothed field is, in all 
likelihood, not a potential field but furnishes the way for the next step.
\item[3.] Take the radial component of this initial smoothed velocity field 
and carry out the line integral along a radial path to obtain the 
velocity potential $\Phi$ at all values of $\vect r$.
\item[4.] Use the potential to derive the vectorial smoothed 
peculiar velocity field by taking the gradient.
\end{description}

Of course there are numerous other considerations that are discussed 
elsewhere(DBF, Newsam et al., \cite{Newsam+93a}, \cite{Newsam+93b}) but 
not crucial to our present discussion.

%
%
\begin{figure*}
  \picplace{7 cm} 
  \caption{Best fit vectorial velocity at given point P. In (a), galaxies A and B lie on a ray. Redshifts provide no information on transverse component of
velocity at P. In (b), galaxies lie on an arc. Both transverse and radial 
components are determined.
\label{fig:nonrad}}
\end{figure*}
%
%
%
\begin{figure}
  \picplace{9.2 cm} 
  \caption{%
The optimal path minimises the error of the velocity potential, and will be
pulled towards regions of high galaxy number density.\label{fig:OptPath}}
\end{figure}
The smoothing procedure in step 2 furnishes us with a 
{\em vectorial\ }velocity field. DBF write down and use only the 
radial component of this
field. Of course there are reasons to suppose the 
radial component will be
more accurately determined than the transverse ($\theta, \phi$) components 
since
the redshifts are essentially telling us the radial components. This can 
be easily seen from simplified picture given in Fig.~\ref{fig:nonrad} {\em (a)},
{\em (b)} where only two galaxies are considered. In case {\em (a)} 
where the observed galaxies are positioned along a ray it is evident that 
the radial component of the initial smoothed velocity field is well 
determined, but there is no information regarding the transverse component. 
Hence the error on the transverse component will be large.
In case {\em (b)}, where the galaxies are transversally 
positioned, the transverse as well as radial component will be reasonably 
well determined. The larger the angle, ${\alpha}$, between the lines of sight 
to the two galaxies, the better determined will be the transverse component,
$v_\theta$. However, even
if the error on the radial component were a factor of ten smaller than on the
transverse, it is still possible to gain advantage by taking non-radial paths.
Fig.~\ref{fig:OptPath} schematically depicts such a situation.

\section{Errors on the Radial and Transverse Components}\label{sec:nonrad}

DBF obtain the initial smoothed velocity field, $\vect v(\vect r)$, at 
arbitrary spatial point $\vect r$, by minimising
\be
{\sum_{i=1}^n (\vect v.\vect {e}_r(\vect {\mathbf{r}_i})-u_i)^2 W(\vect r,\vect {\mathbf{r}_i})}
\ee
where $\vect {\mathbf{r}_i}$ is the position vector of the $i^{th}$ galaxy, and there are 
a total of $n$ galaxies in the survey. $\vect {e}_r(\vect {\mathbf{r}_i})$ is the unit 
vector in the radial direction to the $i^{th}$ galaxy at position 
$\vect {\mathbf{r}_i}$. 
$W(\vect r,\vect {\mathbf{r}_i})$ is the weighting or window function that determines the
relative importance of the $i^{th}$ galaxy. DBF discuss various possible 
window functions, and adopt the one that appears to produce the least bias in their 
simulations. The minimisation is carried out
with respect to the components of $\vect v$ at a chosen position $\vect r$.

Writing
\be{
\vect v(\vect r)=v_r \vect {e}_r(\vect {r})+
v_\theta \vect {e}_\theta (\vect {r})+
v_\phi \vect {e}_\phi (\vect {r})}
\ee
\begin{eqnarray}
s^i_{r r} & = & \vect {e}_r(\vect {r}).\vect {e}_r(\vect {\mathbf{r}_i})\\
s^i_{\phi r} & = & \vect {e}_\phi (\vect {r}).\vect {e}_r(\vect {\mathbf{r}_i})\\ 
s^i_{\theta r} & = & \vect {e}_\theta (\vect {r}).\vect {e}_r(\vect {\mathbf{r}_i})
\end{eqnarray} 
and for expediency  $W^i=W(\vect r,\vect {\mathbf{r}_i})$,
minimisation yields
\begin{eqnarray}
&&\sum_i W^i 
\pmatrix{ s^i_{rr}s^i_{rr}&s^i_{rr}s^i_{\theta r}&s^i_{rr}s^i_{\phi r}\cr
s^i_{rr}s^i_{\theta r}&s^i_{\theta r}s^i_{\theta r}&s^i_{\theta r}s^i_{\phi r}
\cr
s^i_{rr}s^i_{\phi r}&s^i_{\theta r}s^i_{\phi r}&s^i_{\phi r}s^i_{\phi r}\cr}
\pmatrix{v_r\cr v_\theta \cr v_\phi \cr}\nonumber\\
&& = \sum_i \pmatrix{u_i s^i_{rr}W^i\cr u_i s^i_{\theta r}
W^i\cr u_i s^i_{\phi r}W^i\cr}
\end{eqnarray}
which can be written
\be{
\mathbf{ A}\mathbf{ V}= \mathbf{ b}}
\ee
in accordance with the notation of DBF (appendix A). $\mathbf{ V}$ is 
simply the vector of components of the smoothed velocity.

Evidently the inversion of this equation yields all three components of the
initial smoothed velocity field. It should be noted however, as we remarked 
above, that we should expect that the radial component would be normally 
better determined than the transverse components, especially at large distances
or small smoothing lengths.

Since the estimated distances of galaxies are subject to error there will be 
a corresponding error on $\vect v(\vect r)$. Let us write the initial smooth
peculiar velocity field obtained from one realisation as
\be{
\hat {\vect v} (\vect r)= \vect v(\vect r)+ \delta \hat { \vect v}(\vect r)}
\ee
where the hat indicates the estimated velocity. Here we shall assume that all
biases in the estimated velocity have been removed, so that on average, over 
many realisations, the estimated velocity yields the smoothed velocity 
field achieved when no noise is present, but using the same window function 
and redshifts. 
(There is of course no guarantee that the smoothed velocity field obtained 
from the redshifts and window function when no errors are present adequately
represents the true velocity field, but we shall assume that the window 
function has been chosen to minimise this effect, termed the 
sampling gradient bias by DBF).
 
In their analysis, DBF (see their appendix A) carry out a linear error 
analysis in which they derive a bias and
variance on the radial component $v_r$ of $\vect v (\vect r)$ at an arbitrary 
spatial point. Of course the errors on the 
estimated components $\hat v_r (\vect r)$ ,$\hat v_\phi (\vect r)$, 
and $\hat v_\theta (\vect r)$ are
not statistically independent. 

Furthermore, the errors on the estimated velocity at different
spatial points will also be correlated, typically over a distance scale 
determined by the diameter of the window function. This analysis 
can easily be generalised to the vectorial case. 
Since we have assumed that the bias has been removed 
from $\hat v_r(\vect r)$ so that $E(\delta \hat v_r(\vect r))=0$, we can write 
the covariance as
\be{
E(\delta \hat v_i(\vect r)\delta \hat v_j(\vect s))=R_{ij}(\vect r, \vect s)}
\ee
%
where $E$ denotes expected value.
The velocity error autocorrelation function, $R_{ij}(\vect r, \vect s)$, will 
depend on the window function, number density, the dispersion of the distance estimator and
also to some extent on the input peculiar velocities. Our main purpose in 
this paper is to demonstrate the viability of our method and we shall not 
attempt to model accurately $R_{ij}(\vect r, \vect s)$.

\section{Optimal Paths}\label{sec:optpath}

The estimated potential, $\Phi (\vect r)$, 
of the peculiar velocity field is given by the path integral 
\be{
\hat \Phi (\vect r) = -\oint_0^{\vect r} \vect v({\vect s}).d\vect s}
\label{eq:Potent}
\ee
It is important to note that the error, $\delta \hat \Phi (\vect r)$ arises 
only from the
error, $\delta \hat {\vect v}(\vect s)$ in the estimated initial peculiar 
velocity. The radius vector of the path is not a statistical variable,
but is, once the path has been chosen, strictly determined. 
The error on the potential is therefore given by
\be{
\delta \hat {\Phi} (\vect r) 
= -\oint_0^{\vect r}\delta \hat {\vect v}(\vect s).d\vect s}
\ee
The optimal path for obtaining the potential of the 
velocity field at position $\vect r$ we shall
take to be that path for which the variance of the line integral is 
minimised.

If we assume that $\hat {\vect v}$ is unbiased so that 
$E(\delta \hat {\vect v})=0$ at every spatial point, $\vect s$, 
then evidently 
\be{
E(\delta \hat \Phi)=0}
\ee
The variance of $\delta \hat \Phi$ is given by
\begin{eqnarray}
E(\delta \hat {\Phi} (\vect r))^2
& = & E \left ({\oint_0^{\vect r}\delta \hat {\vect v}(\vect s).d\vect s
\oint_0^{\vect r}\delta \hat {\vect v}(\vect t).d\vect t }\right )
\nonumber\\
& = & \int_0^{\alpha}\int_0^{\alpha} R_{ij}(\vect s (\mu ),\vect t (\nu )){x^i}'(\mu )
{x^j}'(\nu ) d\mu d\nu\label{eq:R_ij1}
\end{eqnarray}
Thus the optimal path will be given by
\be{
\delta \int_0^{\alpha}\int_0^{\alpha} R_{ij}(\vect s (\mu ),\vect t (\nu )){x^i}'
(\mu ){x^j}'(\nu ) d\mu d\nu=0}
\label{eq:Var}
\ee
where the path has been parametrised
\be{
x^i = x^i(\mu),\quad 0\le \mu \le \alpha
}\ee

In the case where the autocorrelation function is a delta function,
ie 
\be{
R_{ij}(\vect s, \vect t)=\delta^3 (\vect s -\vect t)\sigma_{ij}(\vect s)}
\label{eq:R_ij2}
\ee
Eq.~(\ref{eq:Var}) simply defines a geodesic on a riemannian space with 
$\sigma_{ij}$ as its metric tensor. Generally, however, we can expect 
components of the initial smoothed field at different spatial points to be 
correlated, and the
correlation length to be of the same order of magnitude as the effective 
radius of the window function.

Although Eq.~(\ref{eq:Var}) is very interesting from a mathematical 
viewpoint, we
shall not proceed further along those lines. In practice we do not know the
form of the autocorrelation function, and it could be best approximated by
numerical simulations. A more natural way to proceed is to use finite element
methods, which we now discuss.

\section{Dijkstra's Algorithm}\label{sec:dijkstra}

We wish to calculate the `best' velocity 
potential, from which the peculiar velocity field may be 
obtained by taking the gradient. This requires determining the potential
$\Phi $ at regular grid points, at least in regions of space where the 
galaxies are sufficiently dense for the reconstruction to be meaningful. 

Suppose we have $N$ gridpoints at which we wish to evaluate the 
potential of the velocity field. Let us assume that the error in moving 
between
the $a^{th}$ and the $b^{th}$ gridpoint is well defined and known for all
$a$ and $b$. This `error length', 
which we shall call an arcweight, should
depend on the number density of galaxies in the joint 
neighbourhood of both gridpoints,
the distance between the gridpoints and the distance of both from the origin.
We take the potential to be zero at the first gridpoint (our
galaxy), and so wish to find the path along which the total error is least.
At first sight it might appear that this is an {\em NP-complete\ }problem
(i.\null e. only soluble exponentially and therefore impractical on even
the largest computers. For a fuller description NP-complete and related problems,
see the reference below).
Luckily this is not the case. Dijkstra's algorithm 
(cf Papadimitriou and Steiglitz, \cite{Papad+82}) finds the exact and global 
solution to this problem in a number of steps that is bounded above by $N^2$. 
This algorithm is one of a class of {\sc max-flow} algorithms used in 
network theory.

\subsection{Description of Dijkstra's Algorithm}

%
%
\begin{figure}
  \picplace{8 cm} 
  \caption{%
Shortest path from O to P, indicated by arrows, can be obtained using
Dijkstra's algorithm.\label{fig:ArcLen}}
\end{figure}
Figure \ref{fig:ArcLen} gives a schematic representations of a network of nodes 
(gridpoints). We write the set of nodes as 
\be{
\mathbf{ N}=\{n_1, n_2, n_3, . . . n_N\}}
\ee
The arcweight between each pair of nodes has been calculated according to some
prescription. Let the arcweight between node $n_a$ and node $n_b$ be $c_{ab}$. 
We shall assume that $c_{ab}=c_{ba}$, although this is not strictly necessary 
for what follows. A path will be determined by its nodes. 
Thus the path $n_{a_1}n_{a_2}n_{a_3}n_{a_4}...  n_{a_M}$ has $M$ nodes 
and total arcweight or pathlength 
\be{
c_{a_1a_2a_3..a_M}=c_{a_1a_2}+c_{a_2a_3}+. . .c_{a_{M-1}a_M}}
\ee 
Clearly we may also write
\be{
c_{a_1a_2a_3..a_M}=c_{a_1a_2a_3 ...a_{M-1}}+c_{a_{M-1}a_M}}
\ee
The problem is to find the path from node $n_1$ to 
every other node $n_a$ that minimises the sum of arcweights. We proceed as
follows.

%
%
\begin{figure}
  \picplace{6.5 cm} 
  \caption{%
Illustration of Dijkstra's algorithm. $\mathbf{M}$ is set of nodes for which 
optimal paths from $m_1$ have been found. $\mathbf{N}$ is the set of all nodes
and $\mathbf{P} = \mathbf{N} - \mathbf{M}$.\label{fig:Dijkstra}}
\end{figure}
Suppose that at some stage we have a set \mathbf{M} of nodes to which 
we have established the minimum pathlength. Let the remaining nodes form 
the set $\mathbf{P}$ (see Fig.~\ref{fig:Dijkstra}). 
Define $s_{\mathbf{M}}(p)$ to be the shortest pathlength from $n_1$ to $p$ 
{\em that uses only intermediate nodes\ }in $\mathbf{M}$. 
This must also be the shortest path from $n_1$ to $p$. Now choose from 
the set of nodes
$\mathbf{P}$ the node $p$ for which the path length is shortest, and add 
it to the set $\mathbf{M}$ and repeat until until the set $\mathbf{M}$ 
contains all the nodes. This algorithm is similar, both in principle and
in implementation, to the {\em Minimal Spanning Tree}, already used
in statistical analyses of large scale structure 
(cf Barrow et al., \cite{Barrow+85})

\section{Application of Dijkstra's algorithm to {\sc Potent} and Results}\label{sec:dijkandPot}

Our main problem in applying this method to {\sc Potent} is how to 
establish the
arcweights between nodes. The arguments we present now are largely heuristic.
Errors between two nodes (gridpoints) we expect to be determined largely by 
the number
density of galaxies in the mutual neighbourhood of the two nodes. The higher 
the galaxy number density the smaller the error. Radial components of the
initial peculiar velocity field will probably have lower errors than the
transverse components, but we might expect the two transverse 
components to have the same errors. If we take the gridpoints $n$ and $m$ 
to be separated by more than the correlation length
of the autocorrelation function, then we can assume that
$\delta \vect v(m)$ and $\delta \vect v(n)$ are uncorrelated.
Hence it would be reasonable to take
\begin{eqnarray}
E(\delta \hat {\Phi} (\vect r))^2 
& = & \sum  E(\delta \vect v(ab).\Delta \vect x(ab))^2 \nonumber\\
& = & \sum E(\delta v_i(ab) \delta v_j(ab))\Delta x^i(ab) \Delta x^j(ab)
\end{eqnarray}
where $n_a$ and $n_b$ are consecutive nodes along the path of integration,
$\Delta \vect x(ab)$ the separation between these two gridpoints,
and $\delta v_i(ab)$ is the error in the $i^{th}$ component of the initial 
smoothed peculiar velocity evaluated at the midpoint of the segment.
Since the variance of the distance estimator 
increases with radial distance squared, we shall take the arcweight
to also scale with $r^2$. To simplify, we disallow arcs between gridpoints 
more than three gridlengths away, and assume that 
\be{
\sigma_{r \theta}=\sigma_{r \phi} = \sigma_{\phi\theta}=0  \hbox{  and }  
\sigma_{r r}=k^2\sigma_{\theta\theta}=k^2\sigma_{\phi \phi}}
\ee
where $k$ is some parameter.

Thus we shall take $c_{ab}$ to be of the form
\be
{c_{ab}=n^\alpha r^2 ((\Delta r)^2+k^2 r^2 (\sin^2 \theta  (\Delta \phi)^2
	  + (\Delta \theta)^2)) }\label{eq:ArcWeight}
\ee
The value of $k$ in the above equation essentially tells one the errors on the
transverse components of the initial field compared with the  radial. This
will obviously depend on the window function. It will also
depend on the actual peculiar velocity field. Very rough simulations indicate
that for simple peculiar velocity fields the transverse components will be
poorly recovered, and consequently $k$ will be large, typically between 5 and
10. $n^\alpha$ is used to used to 
control the number of steps in any path. In general, $\alpha$ can be varied
from $0$ (no preference) to about $-9$ (large numbers of steps strongly
preferred).

%
%
\begin{figure*}
  \picplace{6.5 cm} 
  \caption{%
Some optimal paths with number density of galaxies indicated by dots. Paths in 
figure (a) have low density weighting and most are almost radial. In figure (b),
paths have high density weighting and are non-radial.\label{fig:Paths}}
\end{figure*}
%
%
%
\begin{figure*}
  \picplace{16.5 cm} 
  \caption{%
Potential velocities derived from optimal paths obtained for different density
weightings ($\alpha$) and ratios of transverse to radial errors ($k$). The solid
arrows are the velocities for $\alpha=0$ and $k=9$. Dashed arrows for
$\alpha=-4$ and $k=5$. Galaxy number density is projected onto the plane.
\label{fig:Results}}
\end{figure*}
We have taken a spatial distribution of galaxies to be given by the 
Mathewson redshift survey (Mathewson et al., \cite{Mathew+92}) combined 
with the Burstein Mark II compilation (Burstein et al., \cite{Burstein+87}).
Two peculiar velocity fields are taken corresponding to quiet Hubble 
flow and uniform streaming. Galaxy distances are subjected to 
distance errors, and the
minimum length paths found using arcweights of the form 
in Eq.~(\ref{eq:ArcWeight}) to 
define
the line integral in Eq.~(\ref{eq:Potent}), and hence the velocity potential. 
Figures \ref{fig:Paths} and \ref{fig:Results}
shows the optimal paths for two different heuristic arcweight functions,
and their corresponding rederived velocity fields.

For both fields it turns out that the optimal paths are almost radial. This is
not very surprising since the smoothing uses such a wide window function that
variations in galaxy number density have little effect on causing the optimal path
to deviate from the radial. 
By insisting on arcweights that are heavily dependent on density, and
for which $k \sim 1$ one can achieve non-radial paths. However, in these cases
the recovered potential velocity field is noisy and bears little resemblance
to the input field.

\section{Conclusion and Discussion}\label{sec:conclusion}

For a gaussian window function of radius $1200$kms$^{-1}$, used here as
in conventional {\sc Potent},
the density inhomogeneities are not great enough to produce optimal paths
that are highly non-radial when realistic arcweight functions are chosen.
Although in regions where the data is dense paths deviating from radial
do produce potential velocity fields in agreement with the radial paths
there seems little advantage can be obtained in this way. The reasons for
this somewhat disappointing result lies primarily with the large 
radius of the window function which tends to smooth out the effect of 
number density inhomogeneities. For slowly varying peculiar velocity
fields the transverse components of the initial smoothed peculiar velocity 
obtained from using the window function will be poorly determined, except
perhaps near to our own galaxy, where in any case the recovery from radial
paths is good. Improvements in the derived potential velocity could be
made by integrating along differentiable curves rather than along rectilinear
line segments, but probably the gains would not be commensurate with the
effort involved.

One way to improve on the derived potential would be to 
take the {\em potential\ }at each gridpoint averaged over many paths, which
could be weighted according to their total arcweight. 
The main weakness
of {\sc Potent} seems to us largely to stem from the attributes of the
window 
function. {\sc Potent} does have the advantage over other methods in that 
it is non-parametric. However given the small number of observable galaxies
available for the construction of the velocity field, one is virtually 
obliged
to smooth the velocity field over a very large scale, so that the number of
`independent' values of the inferred smoothed velocity field is small. 
The notion of using an ensemble of paths which are not necessarily 
radial to obtain an ensemble averaged potential is one that can transcend the
use of ad hoc window functions, and possibly overcome the attendant 
difficulties
of sampling gradient and Malmquist bias. 

A natural way to proceed would
be to take an initial potential field obtained by extrapolating the potential
field derived
simply from the peculiar velocities determined at the positions of the 
galaxies in the sample.
To any given point one could then obtain a potential by taking an ensemble 
average of line integrals, weighting the contribution of each one 
according to the total error along its 
path. The situation is reminiscent of Feynman's path integral  
approach to quantum mechanics, where one sums terms of the form
$\exp{(-S)}$ where $S= \int L dt$ is the action, and L the lagrangian. 
The parallel
extends further, as can be seen from Eqns.~(\ref{eq:R_ij1}) and 
(\ref{eq:R_ij2}) where 
the covariance $R_{ij}$
is analogous to the lagrangian. The level of smoothing required could be 
fixed by introducing a `smoothing' parameter, $\lambda$, so that paths 
would be weighted
according to the value of $\exp (- \lambda S)$
We are currently investigating the application of this method.

\section{Acknowledgements}

The authors would like to thank Dr.~Lewis Mackenzie of Glasgow University 
Department of Computing Science for drawing Dijkstra's algorithm to their
attention.
M.\null A.\null H. was funded during this research by an SERC Postdoctoral 
Fellowship.
A.\null N. was funded by SERC Postgraduate Studentship 91306795.


\begin{thebibliography}{}

\bibitem[1985]{Barrow+85}
Barrow, J. D., Bhavsar, S. P. and Sonoda, D. H. 1985,
{M. N. R. A. S.}, {216}, 17

\bibitem[1989]{Bert+89}
Bertschinger, E. and Dekel, A. 1989,
{ Astrophys. J.}, {336}, L5

\bibitem[1990]{Bert+90}
Bertschinger, E., Dekel, A., Faber, S., Dressler, A., and Burstein, D. 1990,
{Astrophys. J.}, {364}, 370

\bibitem[1987]{Burstein+87}
Burstein, D., Davies, R., Dressler, A., Faber, S., Stone, R., Lynden-Bell, D.,
  Terlevich, R., Wegner, G. 1987,
{ Astrophys. J. Suppl.}, {64}, 601

\bibitem[1990]{Dekel+90}
Dekel, A., Bertschinger, E., Faber, S. 1990,
{Astrophys. J.}, {364}, 349

\bibitem[1992]{Dekel+93}
Dekel, A., Bertschinger, E., Yahil, A., Strauss, M., Davis, M., Huchra, J.
  1992,
{Astrophys. J.}, {412}, 1

\bibitem[1992]{Mathew+92}
Mathewson, D., Ford, V., Buchhorn, M. 1992,
{Astrophys. J.}, {389}, L5

\bibitem[1993]{Newsam+93a}
Newsam, A., Simmons, J., Hendry, M. 1993,
in {Proceedings of the 9th IAP meeting: ``Cosmic Velocity
  Fields''} eds Lachi\`{e}ze-Rey, M. and Bouchet, F.,
(Editions Fronti\`{e}res)

\bibitem[1993]{Newsam+93b}
Newsam, A., Simmons, J., Hendry, M. 1993,
{Bias minimisation in {\sc Potent}},
{Astron. Astrophys.}, 
Submitted

\bibitem[1982]{Papad+82}
Papadimitriou, C. and Steiglitz, K. 1982,
{Combinatorial Optimisation -- Algorithms and Complexity},
(Prentice Hall)
\end{thebibliography}
\end{document}